# At the threshold of distributed Kerr-Lens mode-locking in a Cr:ZnS waveguide laser

**MAKSIM DEMESH,**[1*] **VLADIMIR L. KALASHNIKOV,**[1] **EVGENI SOROKIN,**[2,3] **NATALI GUSAKOVA,**[1] **ALEXANDER RUDENKOV,**[1] **AND IRINA T. SOROKINA**[1,2]

[1] *Department of Physics, Norwegian University of Science and Technology, N-7491 Trondheim, Norway*
[2] *ATLA lasers AS, Richard Birkelands vei 2B, 7034 Trondheim Norway*
[3] *Institut für Photonik, TU Wien, Gußhausstraße 27/387, A-1040 Vienna, Austria*

*\*maksim.demesh@ntnu.no*

**Abstract:** We demonstrate feasibility of spatiotemporal mode-locking in a mid-IR Cr:ZnS waveguide laser based on the nonlinear spatial mode coupling. The experiment shows efficient power scaling with the excitation of intra-mode beatings, causing a pronounced Q-switching which could cause a nonlinear mode-coupling. We suggest that a particularly high nonlinearity in Cr:ZnS combined with multimode waveguide leads to a soft aperture induced by a pump beam. The latter provides an effective spatial mode control in a nonlinear multimode waveguide and, thereby, opening the way to the birth of a spatiotemporal dissipative soliton, or light bullet, formation. Thus, forming the basis for the future distributed Kerr-Lens Mode Locking in the energy-scalable solid-state waveguide (or yet unrealized fiber) laser.

## 1. Introduction

The modern trend in generating and exploring laser *dissipative solitons* (DS) [1] is based on a controllable enhancement of self-organizing effects induced by nonlinearities and affected by both temporal and spatial degrees of freedom [2]. The underlying concept involves using a well-structured and enhanced Kerr-nonlinearity in a laser cavity. A partial realization of this concept is a *distributed Kerr-lens mode-locking* (DKLM), for the first time suggested and realized in a so called XX cavity in [3] and recently explored in [4]. The authors of [3] have demonstrated that it is possible to spatially redistribute the enhanced KLM action in an oscillator. In a solid-state laser, this technique used an additional nonlinear medium which affects a laser beam via its self-focusing and, thereby, changes the effective gain in a laser resonator as a whole [5]. That unites a solid-state laser with a full-fiber one, characterized by a distributed nonlinearity, and could provide with realizing the spatiotemporal mode-locking and generation of spatiotemporal DS [6]. A remarkable breakthrough in the energy scalability of such type of oscillators was demonstrated in both anomalous [3,5,7] and normal group-delay dispersion regimes (ADR and NDR, respectively) [8].

Exploring DS in NDR fiber lasers [9] opened new directions in exploiting the nonlinearities for the physics of well-localized, coherent light structures and bridged fiber and solid-state ultrafast laser photonics [2]. A significant difference is that fibers, being essentially waveguides, allow nonlinear propagation of many interacting spatial modes (so-called multimode fibers, MMF) [10]. As a result, a field can be composed of a multitude of spatial modes, which could be coupled through a nonlinear process, i.e., nonlinear refraction characterized by the coefficient $n_2$. Since waveguiding results from the action of "confining potential" defined by spatially-dependent refractive index $n(x,y)$ in "graded refractive index fibers," (GRINs), and nonlinear lattices [11,12], the higher-order modes may relax to the ground-state, i.e., lowest-order spatial mode. That is a process of the so-called *mode-cleaning* (condensation of spatial modes), which can form a *spatial soliton* [13-15]. One has to emphasize that a trapping potential is vital for multidimensional soliton stabilization [14,16, 17].

The paper presents an attempt to connect this strategy underlying a nonlinear mode-condensation in an MMF (as well as a waveguide, photonic-crystal fiber, and an array of waveguides) with a DKLM technique that could allow a *spatiotemporal mode-locking* (STML) [6,18]. We suggest that using dissipative (i.e., with a nozero imaginary part of refractive index $\Im[n(x,y)] \neq 0$) trapping potential would provide a self-sufficient mode-locking mechanism and, as a consequence, spatiotemporal DS (STDS) generation without involvement of other nonlinear dissipative processes. Often, the mode-locking requires some "supplementary" dissipative nonlinearity, like loss saturation in a semiconductor absorber, nonlinear polarization rotator, etc. [19-21]. Here, we do not use "external" nonlinear gain/loss mechanisms but explore the potential of the passive mode-locking inside the multimode waveguide. Feasibility of the approach using transversely graded complex trapping potential was demonstrated theoretically in [22] and is in line with stabilizing action of the localized gain/loss on a spatial DS [23]. In this work, we explore feasibility of this approach in a multimode Cr:ZnS waveguide laser.

Summarizing, the paper extends our earlier concept of DKLM, uniting a waveguide transversely graded amplification, KLM, and DS formation processes under well-controllable spatial multi-mode guiding conditions. This broader DKLM concept covers the STML all-fiber and solid-state oscillators, where our crystalline Cr:ZnS waveguide presents a good "all-in-one" model for the combined action of all these processes.

## 2. Experimental

The fabrication of a depressed cladding channel waveguide in a 7 mm $Cr^{2+}$:ZnS single crystal with a doping concentration of $1.4 \times 10^{18}$ cm$^{-3}$ was performed by ultrafast laser inscription (see Table I) [24], that allowed us obtaining waveguides with single-pass losses of approximately 0.6 dB/cm and refractive index change $\Delta n$ of $3.6 \times 10^{-3}$. The core diameter of a waveguide that was used in the present experiment was approximately 50 μm, in which multimode propagation of radiation could be supported (see Fig. 1).

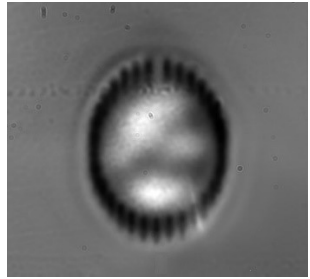

Fig. 1. Multimode propagation of 1.55 μm laser radiation in the 50 μm waveguide.

The $Cr^{2+}$:ZnS sample with inscribed waveguide was mounted on a passive copper heatsink and placed into the microchip laser cavity (see Fig. 2). The plane input mirror (IM) was highly transparent (>99.5%) for the pump wavelength and highly reflective (>99.5%) for 2.1–2.5 μm. The plane output coupler (OC) has a transmission of 40% at the laser wavelength. The pump beam from *IPG Photonics* Er:fiber laser at 1610 nm was focused into a spot of ~50 μm diameter inside the waveguide. To avoid the back-reflection from the uncoated crystal facet to the pump laser, we used a Faraday isolator (FI).

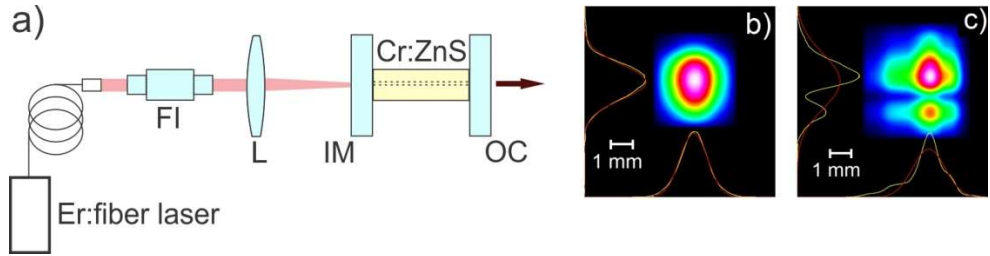

Fig. 2. Schematic of the Cr:ZnS waveguide laser (a) and its output beam profile in single- (b) and multi-mode (c) operation.

At an incident pump power of 4 W we obtained output power of 240 mW with a slope efficiency of 7.4% (Fig. 3). The emission spectrum in the range of 2277–2287 nm confirms the multimode operation. With a smaller OC transmission of 10%, the waveguide was unexpectedly damaged at an output power of only about 250 mW, corresponding to approximately 2.5 W of an intracavity power – a much too low power level for CW beam to incur any type of damage in such a robust crystal as ZnS. Indeed, the optical damage threshold of ZnS is the highest among semiconductors [25] enabling tens of Watts of output power in CW regime at >100 W of pump power [26, 27].

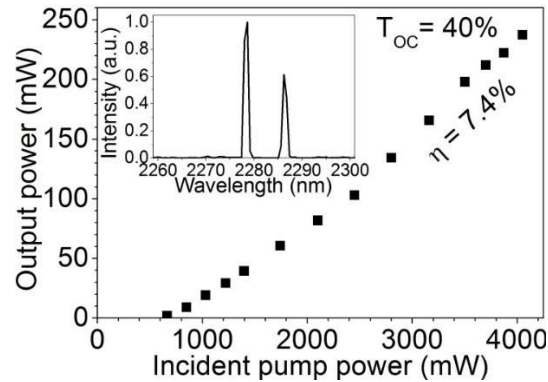

Fig. 3. Laser output characteristic of the waveguide Cr:ZnS laser. Inset - laser spectrum

The damage was more likely caused by switching between the CW and mode-locked regimes. Such a transient process has been reported in Refs [28, 29]. We suppose that the high material nonlinearity in a Cr:ZnS waveguide leads to the easy formation of both a single mode (Fig. 2(b)) as well as multi-mode (Fig. 2(c)) regimes. Interestingly, as we show in the Section 3.2, the associated nonlinear spatiotemporal dynamics could result in DKLM in a waveguide laser frequently accompanied by Q-switch-mode-locking with strong pulse bursts. The latter was likely the phenomenon observed in our waveguide, which acquired damage when reaching the threshold of DKLM. To check this, we carried out numerical analysis using close to experimental laser parameters.

## 3. Spatial mode control in a "soft-aperture" waveguide laser

For modeling, let us consider a waveguide laser composed of a crystalline active waveguide enclosed in the Fabri-Perót cavity and driven by a Gaussian pump beam. The main idea is to use a Gaussian pump beam as a "soft aperture" guiding laser beam and controlling ("cleaning") its mode structure so that field condensates into lowest-order mode.

## 3.1 Model

We begin with a distributed model describing the evolution of the $(x,y)$-dependent field $a(x,y,z)$ along the $z$-axis of a waveguide consisting of an active medium with the linear $n(x,y)$ and nonlinear $n_2$ refractive indexes, respectively, and placed within a resonator with sufficiently high Q-factor $\propto L^{-1}$ ($L$ is a resonator net-loss coefficient). Under the action of $(x,y)$-graded pump beam, which is described by a saturable gain coefficient $G(z,x,y)$, the evolution can be described by a kind of the driven nonlinear Helmholtz equation:

$$\left[\frac{1}{2k_0 n_0}\left(\Delta_{x,y} - \frac{\partial^2}{\partial z^2}\right) + \frac{k_0}{2n_0}(n(x,y)^2 - n_0^2)) + i(L - G(z,x,y)) + k_0 n_2 |a(z,x,y)|^2 + i\frac{\partial}{\partial z}\right] a(z,x,y) = 0, \quad (1)$$

where $\Delta_{x,y}$ is a Laplace operator in the Cartesian coordinates, $n_0$ is a waveguide cladding refractive index, $k_0$ is a free-space wavenumber, and the second-order derivative over $z$ contributes only for large beam numerical apertures NA [30]. The terms in Eq. 1 define: (I) – diffraction, (II) – trapping potential induced by GRIN, (III) – transversely graded saturable gain with the coefficient $G$ and linear net-loss $L$ averaged over a resonator round-trip, and (IV) – Kerr-nonlinearity.

For a low NA, Eq. (1) can be reduced to a kind of the nonlinear driven Schrödinger equation, or the Lugiato-Lefever equation (LLE), which is broadly used as a model of spatial and temporal soliton and pattern formation in driven Kerr resonators, micro-cavity and fiber lasers, VCSELs, etc. [31-33].

Using the ansatz $\psi = a\exp(-iV_0 z)$ (here $V_0 = (k_0/2n_0)(n_1^2 - n_0^2)$, $n_1$ is a waveguide core refractive index), the dimensionless mean-field LLE can be written as [34-36]:

$$\left[\frac{1}{2}\Delta_{X,Y} - (X^2 + Y^2) + i(\Lambda - G(Z,X,Y)) + |\psi(Z,X,Y)|^2 + i\frac{\partial}{\partial Z}\right]\psi(Z,X,Y) = 0, \quad (2)$$

where the following rescalings and normalizations are used [37]: $(X,Y) = (x,y)/w_T$. Here $w_T = \sqrt[4]{w_0^2/k_0^2 n_0 \delta}$ is a transverse coordinate scale, $w_0$ is a size of the parabolic trapping potential induced by GRIN, $\delta = n_1 - n_0$ is a "trapping potential depth", and $(k_0/2n_0)(n_1^2 - n(r)^2)) \simeq k_0(n_1 - n(r))$ is assumed. $Z = z/\zeta$, $\zeta = k_0 n_0 w_T^2$, $\Lambda = LL_w/\zeta$, and the field intensity is normalized as $|\psi|^2 = k_0 n_2 \zeta |a|^2$. $L_w$ is a waveguide length.

As we consider a waveguide laser, a saturable and transversely profiled gain $G(Z,X,Y)$ plays a decisive role. Since the gain relaxation time $T_r$ of the order of microseconds exceeds both characteristic scales $c/\zeta$ and $1/L_w$ ($c$ is the velocity of light), the gain saturates integrally over the characteristics scales [38]:

$$G(Z,X,Y) = \frac{g_0(X,Y)}{1+|\psi(Z,X,Y)|^2/I_s}, \quad (3)$$

$$g_0(X,Y) = (N_{Cr}\sigma_{em}\zeta) \times \left(\frac{W}{\Phi_s}\right) \times \exp(-(X^2+Y^2)/w_p^2), \quad (4)$$

$$W = \sigma_a P/S\varepsilon_p, \quad (5)$$

$$\Phi_s = W + 1/T_r, \quad (6)$$

$$I_s = (\varepsilon\Phi_s/\sigma_{em}) \times (k_0 n_2 \zeta). \quad (7)$$

Here, we assume the Gaussian transverse profile of the gain with the size $w_p$ (4), and no reflection of the pump from the OC. Below, we will represent this transverse profile as a parabolic expansion in the vicinity of the waveguide axis. Such transversely confined nonlinear gain acts as a "soft aperture" with an effective size $d \approx w_p\sqrt{G-L}$. The maximum gain defined by the first multiplier in (4) contributes together with the pump rate $W$ (5) ($P$ is a pump power, $S$ is an effective pump beam area) and the "gain saturation rate" $\Phi_s$ (6). The latter defines the normalized gain saturation intensity $I_s$ (7). The used physical parameters are combined in Tables I and II.

In the case of a soft aperture, when the dissipation is transversally graded not due to the waveguide structure [22] but due to the pump beam transverse profile $g_0(X,Y)$, the trapping potential induced by such radially varying gain can be characterized by a soft aperture "increment" $\rho$, defining the trapping potential depth and profile. The momentum method allows writing such increments (see Eq. (S3.16) in [39]) in the following form:

$$\rho = -\frac{2w^2 w_p}{(w_p^2 + 2w^2)^{3/2}}. \tag{8}$$

Here, we assume Gaussian profiles for pump and laser beams, with the sizes $w_p$ and $w$, respectively. Physically, $\rho$ is a measure of the soft-aperture trapping "strength" so that the depth and profile of the trapping potential (see right inset in Fig. 4) define the effectiveness of the dissipation-induced mode-cleaning. In the quantum-mechanical language, this potential affects the level occupancy ("levels" correspond to the transverse modes). Increase of the ground level (i.e., the fundamental mode) occupancy for $w \approx w_p$ (see left inset in Fig. 4) is equivalent to the mode-cleaning action.

**Table I. Physical parameters of Cr:ZnS waveguide laser [34]**

| Laser wavelength, $\lambda$ | $2.272 \times 10^{-4}$ cm |
|---|---|
| Pump wavelength, $\lambda_p$ | $1.61 \times 10^{-4}$ cm |
| Absorption cross-section, $\sigma_a$ | $10^{-18}$ cm$^2$ |
| Emission cross-section, $\sigma_{em}$ | $1.49 \times 10^{-18}$ cm$^2$ |
| Gain relaxation time, $T_r$ | $4.3 \times 10^{-6}$ s |
| Cr-ions concentration, $N_{Cr}$ | $1.4 \times 10^{18}$ cm$^{-3}$ |
| Core refractive index, $n_1$ | 2.2706 |
| Cladding refractive index, $n_0$ | 2.2629 |
| Nonlinear refractive index, $n_2$ | $9 \times 10^{-15}$ cm$^2$/W |
| Group-delay dispersion coefficient, $\beta_2$ | 9850 fs$^2$/cm |

**Table II. Simulation parameter scales for a Cr:ZnS waveguide laser**

| Waveguide length | 0.8 cm |
|---|---|
| Transverse coordinate unit $w_T$ (10% net-loss, $w_0 = 3 \times 10^{-3}$ cm) | $9 \times 10^{-4}$ cm |
| Effective diffraction length $\zeta$ | 0.022 cm |
| Axis small-signal gain $g_0$ (2 W pump power, $3 \times 10^{-3}$ cm pump beam size) | $4.3 \times 10^{-3}$ |
| Rescaled loss coefficient $\Lambda$ | $2.8 \times 10^{-3}$ |
| Core refractive index, $n_1$ | 2.2706 |
| Cladding refractive index, $n_0$ | 2.2629 |
| Inverse squared gain bandwidth, $\tau$ | 9÷20 fs$^2$ |
| Intensity normalization parameter, $k_0 n_2 \zeta$ | 0.18 TW$^{-1}$cm$^2$ |

## 3.2 Spatial structure and dynamics

Table II shows the parameters appearing in Eq. (1) for our Cr:ZnS waveguide laser. The results of numerical simulations based on these parameters and the model described above are shown in Figs. 4–6.

One may see that the best overlapping of both dissipative and non-dissipative trapping potentials, i.e., $w_0 \approx w_p$ (see the black curve in the right inset in Fig. 4), produces the most "clear" spatial mode with minimal contribution of higher-order modes (left inset in Fig. 4). Simultaneously, the peak power dynamics tends to be stationary (black curve in Fig. 4).

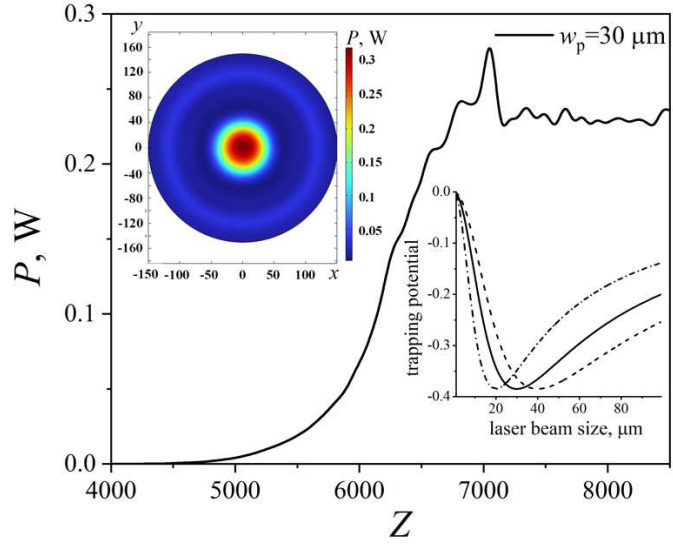

Fig. 4. Evolution of the beam power with $Z$ for $w_p = w_0 = 3 \times 10^{-3}$ cm. Other parameters are shown in Tables I and II. The left inset shows the power profile on the dimensionless $(X, Y)$-plane at $Z = 8500$. The right inset demonstrates the soft-aperture trapping potential $\rho$ on the beam size $w$ for $w_p = 3 \times 10^{-3}$ cm (solid), $4 \times 10^{-3}$ (dashed), and $2 \times 10^{-3}$ (dash-dotted line) cm.

Both "sharper" ($w_p < w_0$, the dash-dotted curve in Fig. 4) or "smoother" ($w_p > w_0$, the dashed curve in Fig. 4) dissipative potentials lead to a Q-switching behavior (e.g., the solid curve in Fig. 5) with excitation of higher-order spatial modes.

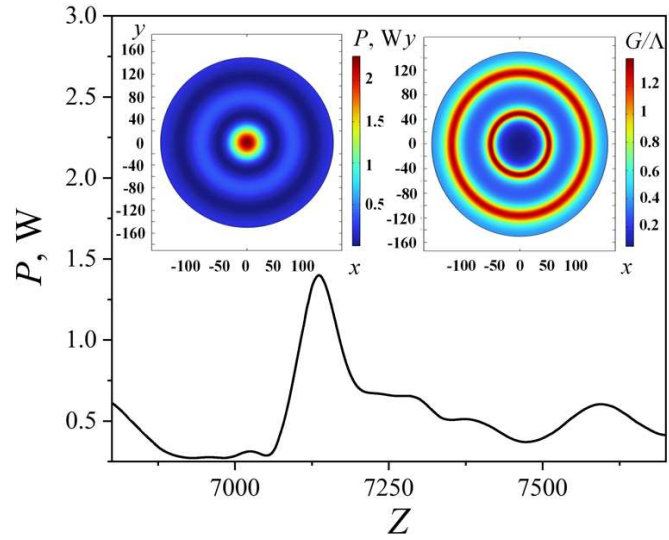

Fig. 5. Evolution of the beam power with Z for $w_0 = 3 \times 10^{-3}$ cm and $w_p = 4 \times 10^{-3}$ cm. Other parameters are shown in Tables I and II. The left inset shows the power profile on the dimensionless $(X, Y)$-plane near the power peak at $Z \approx 7100$. The right inset demonstrates the corresponding relative gain profile. The soft-aperture trapping potential is shown by a dashed line in the right inset in Fig. 4..

Such modes are clearly manifested for $w_p < w_0$ with simultaneous squeezing of ground mode (Fig. 6). This mode-squeezing enhances Q-switching, i.e., enlarges the peak power bursts that could, on one hand, result in unintended damage, or enforce a tendency to DKLM (see next Section).

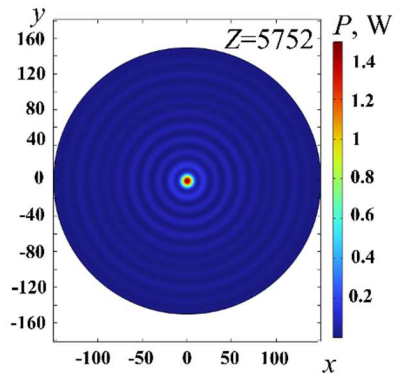

Fig. 6. Power profile on the dimensionless $(X.Y)$-plane at $Z = 5752$ for $w_0 = 3 \times 10^{-3}$ cm and $w_p = 4 \times 10^{-3}$ cm. Other parameters are shown in Tables I and II. The corresponding soft-aperture trapping potential is shown by the dash-dotted line in the right inset in Fig. 4.

The transversely graded gain saturation (right inset in Fig. 5) enhances the spatial modes competition like that for the longitudinal modes. And similarly to the latter, one could pose the task of synchronizing them through a nonlinear process [6,18], i.e., Kerr nonlinearity in our case. The feasibility of such spatio-temporal mode synchronization in a waveguide laser will be considered in the next Section.

## 4. Spatiotemporal mode-locking in a waveguide laser

As it was demonstrated experimentally for a femtosecond solid-state oscillator [4] and predicted for a fiber laser [22], the mode-cleaning mechanism induced by graded dissipation acting as a sort of "soft-aperture" could allow realizing the DKLM and the energy-scalable femtosecond pulse generation without the involvement of the additional nonlinear processes.

The following reasoning helps to obtain the rough estimation of DKLM efficiency [39,41,42]. Let us consider a Gaussian beam propagating in a monolithic nonlinear medium. Such propagation can be described by rescaling the imaginary part of the $q$ beam parameter: $q_0^{-1} = -i\sqrt{1-K}\lambda/\pi w_0^2$, where $w_0$ is the waist size of the input (plane) beam, and $K = P/P_{cr}$ ($P$ is the instant power, $P_{cr} = \alpha \lambda^2/4\pi n_0 n_2 \approx 0.37$ MW is the self-focusing critical power, $\alpha \approx 1.84$ [43]). The rescaled quasi-free-space propagation over the distance $z$ results in a new $q$-parameter $q = q_0 + z$. Then, the imaginary part of $q^{-1}$ should be newly rescaled that gives a new beam size (see Section "Rought estimation of the self-amplitude modulation parameters" in [44]):

$$w^2 = w_0^2 + \frac{(1-K)\lambda^2 z^2}{\pi w_0^2}. \tag{9}$$

If one estimates the loss induced by a hard aperture with the size $D$ as $L = \exp(-D^2/w^2)$, the exponent can be written as:

$$\frac{D^2}{w^2} = \frac{D^2}{w_0^2 + (1-K)\pi^2 \lambda^2/\pi^2 w_0^2}. \tag{10}$$

Now, the loss due to the hard aperture can be expressed under the condition of $\zeta^2/w_T^2 \gg 1$ (far-field approximation) as $L \approx \lambda^2 z^2/\pi^2 w_0^2 D^2 \approx 0.03$ for $w_0 = w_p = D = 30~\mu$m and $z = \zeta$. Under the action of self-focusing, a beam with the power $P$ will "saturate" this loss by $\delta L \sim P \times \lambda^2 z^2/\pi^2 D^2 w_0^2 P_{cr}$, where the multiplier is a so-called inverse "loss saturation power." This power is about 11 MW for the parameters above, which is close to the estimations of [42]. Nevertheless, the above estimation of the saturation power is rough, and one may expect reducing the saturation power down to ~100 kW due to its $w_0^2$ dependence, which is feasible in the solid-state KLM oscillators [40]. The feasibility of the DKLM in a waveguide laser is the aim of our further analysis.

### 4.1 Model and numerical simulations

We use the model based on the LLE (2) but dimensionally extended by adding the "fast-time" (or "local time") coordinate $t$. That allows for describing the pulse dynamics in the co-moving frame $(X, Y, t)$. Since an ultrashort pulse is sensitive to a group-delay dispersion (GDD), it has to be taken into account, which leads to the following dimensionless equation:

$$i\frac{\partial \psi(Z,X,Y,t)}{\partial Z} = -\frac{1}{2}\left[\frac{\partial^2}{\partial X^2} + \frac{\partial^2}{\partial Y^2} \mp \frac{\partial^2}{\partial t^2} - 2i\tau\frac{\partial^2}{\partial t^2}\right]\psi + (X^2 + Y^2)\psi - |\psi|^2\psi - i[\Lambda' + \kappa(X^2 + Y^2)]\psi. \tag{11}$$

The minus/plus signs before $\frac{\partial^2}{\partial t^2}$ correspond to normal and anomalous GDD, respectively.

We assume that a laser operates close to a quasi-steady state in the vicinity of the lasing threshold, where the saturated net-loss coefficient $\Lambda' = \Lambda - G'$ is close to zero [22]. For simplicity, we approximate graded saturated gain as $G'(X,Y) = G'(0,0) \times [1 - (X^2 + Y^2)w_p^{-2}]$, so that $\kappa = G'(0,0)w_p^{-2}$ in Eq. (11). The dimensionless form of Eq. (11) assumes the above normalizations and the normalization of a local time coordinate $t$ to $\sqrt{|\beta_2|\zeta}$.

Another essential modification is adding a spectral dissipation described by the parameter $\tau$ in Eq. (11). This parameter is inversely proportional to the squared spectral width of the gain band or spectral filter (Table II). The resulting equation can be considered as a version of the generalized Gross-Pitaevskii equation, which is a well-known tool for analyzing trapped BEC [45] so that there is a strong link between the dynamics of a laser soliton and the BEC [46].

We assume that a laser operates in the normal GDD regime (Table I), which would lead to the formation of a chirped DS. Such a regime is desirable from the point of view of DS energy scalability [1]. The feasibility of such regime in a fiber laser with graded complex refractive coefficient was conjectured in [22]. Here, let us consider the possible realization of a similar regime in the normal GDD in a waveguide laser, where a pump beam plays a role of soft aperture like that considered above. Like a solid-state laser with DKLM [4], self-focusing, which affects the ensemble of multiple transverse modes, could result in the spatiotemporal mode-locking with the formation of spatiotemporal soliton, a *light bullet* [2].

We solved Eq. (11) with the help of COMSOL software using the finite element method. For reducing the computational time, the axial symmetry was assumed, i.e., a radial coordinate $R = \sqrt{X^2 + Y^2}$ was used with the corresponding Laplacian in (11). The simulations demonstrated the formation of a DS from the initial low-amplitude seed after a long transitional stage ($\gtrsim 20000\ \zeta$).

Some preliminary results are shown in Fig. 7, demonstrating the spatio-temporal profiles of the resulting DS. DS is stretched in the time domain in comparison with their counterparts in the anomalous GDD regime (see Fig. 8 and [22,46,47]), has a visible "truncation" on the temporal edges, and is especially interesting in spatial boundaries, too (Fig. 7, a).

The stretched and truncated temporal profile (Fig. 7, d) testifies the presence of a strong chirp, promoting energy accumulation and stability. A similar pattern appears in the spatial domain as well, which manifests the contribution of the spatial chirp to the energy balance of DS [48]. Thus, one may conjecture that a spatiotemporal mode-locking based on the DKLM mechanism is feasible in a waveguide laser without an additional nonlinear mechanism. As a result, a "slab-like" spatiotemporal DS may develop that demonstrates an unusual example of space-time duality [49] in a nonlinear system far from equilibrium.

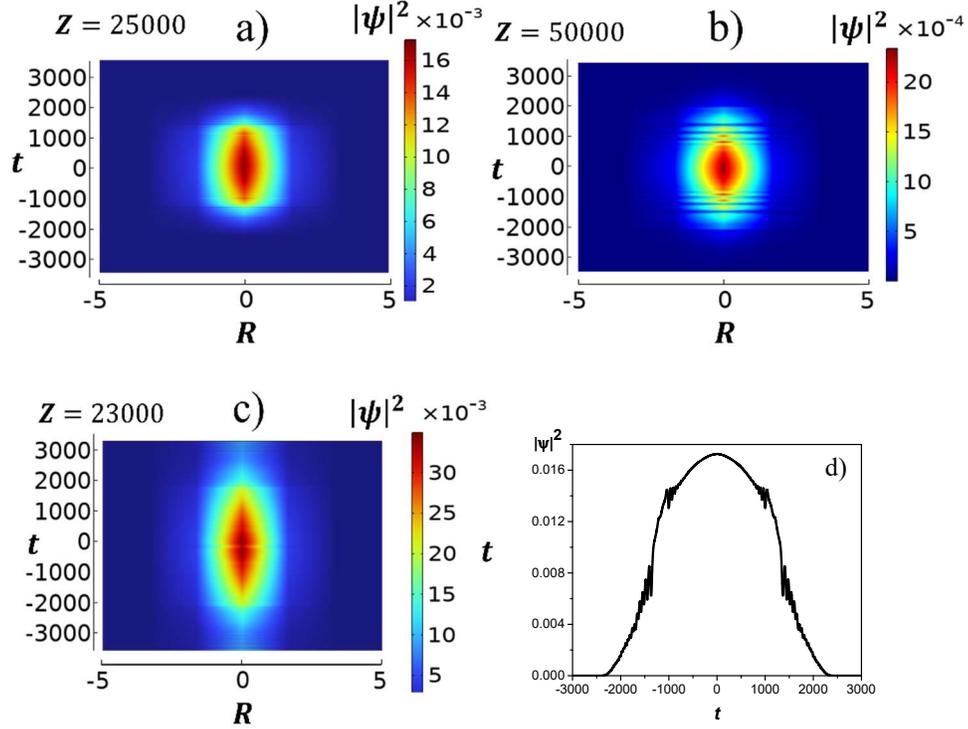

Fig. 7. Spatiotemporal contour plots of the dimesionless DS intensity $|\psi|^2$ on the dimensionless $(R,t)$-plane at the noted distances $Z$ for for the normal GDD $\kappa = 7 \times 10^{-4}$ (a, b), $6 \times 10^{-4}$ (c), $\tau = 10^{-3}$ (a, c), $10^{-2}$ (b), and $\Lambda' = -10^{-3}$. (d) – dimensionless intensity profile on a beam axis for the parameters of (a).

Since DS is chirped, spectral filtering contributes to its spatiotemporal confinement [50]. In the case considered, the DS energy out-flow [48] induces Kolmogorow's cascade to higher wave numbers (in the spatial domain) and frequencies (in time) [51]. The spectral filtering and soft aperture cut them off, enhancing spatiotemporal localization. However, an overgrowth of spectral dissipation causes fragmentation of the DS edges in the time domain (Fig. 7, b) with the subsequent multi-pulse generation.

A decrease of the spatially-graded loss (i.e., broadening of "a soft aperture" with increasing $w_p$, Eq. (8) and inset in Fig. 4) stretches a DS in the time domain (Fig. 7, c), and, for a too broad aperture, leads to a DS breakdown. The latter takes place for a too-narrow aperture, as well.

One should note, that the intracavity DS energy $E$ is $2\pi \iint r|\psi|^2 dr dt \approx 360$ nJ for Figs. 7, a, d. If the intracavity average power equals 5 W, this energy level could be achieved for a laser repetition rate of $\approx 14$ MHz that requires using a stretched resonator like that used in a chirped pulse oscillator [52]. One must note that the energy above could not be considered a precise quantitative estimation because it is $\Lambda'$-dependent. The latter itself depends on energy, and can be estimated only for the specific experimental environment [53].

As was mentioned in [6], the mode-locking self-start from an initial Q-switching is a typical scenario for a chirped-pulse oscillator with enhanced nonlinearity. The main trouble is the possibility of optical damage, as we also observed. The spatiotemporal model based on Eq. (11) does not allow assessing this issue because it neglects the dynamical gain saturation, which is the aim of further studies.

The alternative approach is using an anomalous dispersion for the ST DS generation. That would require a compensation of the waveguide dispersion by, for instance, chirped mirror, as it takes place in usual solid-state lasers. As was pointed out in [29], using the anomalous GDD facilitates mode-locking self-start so that a possible destructive action of Q-switching, fraught with optical damage at the initial stage of pulse formation, can be avoided.

Our calculations demonstrate the feasibility of such a spatiotemporal mode-locking regime. Fig. 8 illustrates the intensity contour plot and its slice (inset) on the beam axis for a spatiotemporal soliton developed under anomalous GDD. A similar regime was considered in detail for a multimode fiber laser in [22].

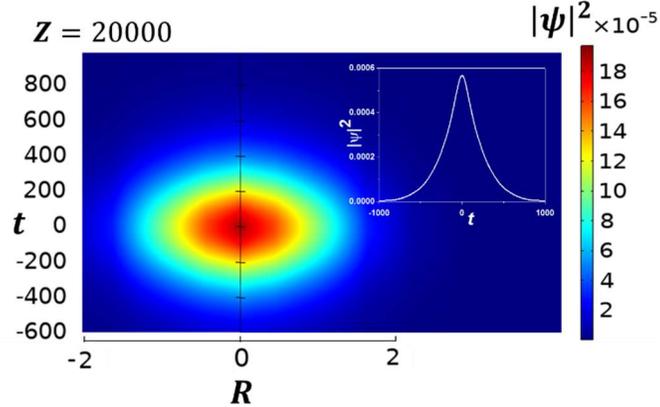

Fig. 8. Spatiotemporal contour plots of the dimesionless DS intensity $|\psi|^2$ on the dimensionless $(R,t)$-plane at the noted distance $Z$ for the anomalous GDD, $\kappa = 7 \times 10^{-4}$, $\tau = 10^{-3}$, and $\Lambda' = -10^{-3}$. Inset – dimensionless intensity profile on a beam axis.

One may see that a pulse is not stretched and has a profile similar to the usual Schrödinger soliton, i.e., it is chirp-free (or almost chirp-free). As a result, it is essentially shorter than the chirped DS (Figs. 7 a,d) and has a smaller peak power and energy ($\approx 10$ nJ for the assumed net-GDD of $-9850$ fs$^2$/cm in Fig. 8).

The 1D-soliton width is related to the dispersion by the scaled soliton area theorem $|\psi_0|T = \sqrt{|\beta_2|}$ [54]. Since the soliton energy $E = 2|\psi_0|^2 T$, there are two alternatives for the energy scaling [55]: either $E \propto \sqrt{|\beta_2|}$ if the peak power is confined from the above (scaling by the pulse width, e.g., a soft-aperture DKLM considered above) or $E \propto |\beta_2|$ if the soliton width is confined from the bottom (scaling by the peak power, e.g., a hard-aperture DKLM like that considered in [22]). The first provides a slower dependence on dispersion, but the latter results in the minimum pulse width confined by $T \approx \sqrt{\tau L/|\rho|}$ [55]. Modification of these simple rules for a 2D-DS needs further study.

## 5. Conclusion

We demonstrated the operation of Cr:ZnS channel waveguide laser with an output power of 240 mW and a slope efficiency of 7.4% at a wavelength 2.28 μm. We suggest that the power scalability and controlled multimode spatial structure of a laser could allow realizing a spatiotemporal soliton due to the distributed Kerr-lens mode-locking under the action of a transversly profiled pump beam.

The multimode interaction may cause a Q-switching operation which is controlled by a soft aperture created by a spatially graded gain due to the pump beam radial profile. One may conjecture that such behavior could be an advantage in the distributed Kerr-lens mode-locking due to the high nonlinearity of ZnS causing mode self-cleaning. Using a waveguide provides the conditions for enhancing the latter process under well-controlled spatial multi-mode guiding conditions.

Numerical simulations demonstrated that the spatiotemporal dissipative soliton develops in a waveguide laser under the action of self-phase modulation, both normal and anomalous group-delay dispersion, spectral and spatially-graded dissipations. The latter results from a Gaussian-profiled pump beam providing a soft-aperture action on a waveguide laser beam. The control of effective soft aperture size and relative contribution of dispersion and spectral dissipation stabilizes a spatiotemporal soliton. That could provide a breakthrough in realizing spatiotemporal mode-locking and open new frontiers in generating high-energy ultrashort pulses from solid-state, waveguide, and fiber oscillators.

## Acknowledgments

The work is supported by the Norwegian Research Council projects #303347 (UNLOCK), #326503 (MIR).

## Disclosures

The authors declare no conflicts of interest